\documentclass{aa}  
\usepackage{graphicx}
\usepackage{txfonts}
\begin{document}
\title{Mapping the large-scale anisotropy in the WMAP data} 

\author{
A. Bernui\inst{1},
B. Mota\inst{2},
M.J. Rebou\c{c}as\inst{2},
\and
R. Tavakol\inst{3}
        }

   \offprints{M.J. Rebou\c{c}as} 

   \institute{Instituto Nacional de Pesquisas Espaciais, 
   Divis\~{a}o de Astrof\'{\i}sica, Av. dos Astronautas 1758, 
   12227-010, S\~ao Jos\'e dos Campos -- SP, Brazil; 
   \email{bernui@das.inpe.br}   
\and
   Centro Brasileiro de Pesquisas F\'{\i}sicas,
   Rua Dr. Xavier Sigaud 150,
   22290-180 Rio de Janeiro -- RJ, Brazil \\       
   \email{brunom@cbpf.br};   
   \email{reboucas@cbpf.br}     
\and
   Astronomy Unit, School of Mathematical Sciences,  
   Queen Mary, University of London, Mile End Road, London E1 4NS, UK; 
   \email{r.tavakol@qmul.ac.uk}
             }

   \date{Received / Accepted}
   
\authorrunning{Bernui, Mota, Rebou\c{c}as \& Tavakol}
\titlerunning{\ Mapping large-scale anisotropy in WMAP data}

\abstract
{}
{Analyses of recent cosmic microwave background (CMB) observations have 
provided increasing hints that there are deviations in the universe 
from statistical isotropy on large scales. 
Given the far reaching consequences of such an anisotropy for our 
understanding of the universe, it is important to employ alternative 
indicators in order to determine whether the reported anisotropy is 
cosmological in origin and, if so, extract information that may 
be helpful for identifying its causes.}
{Here we propose a new directional indicator, based on separation 
histograms of pairs of pixels, which provides a measure of departure 
from statistical isotropy. 
The main advantage of this indicator is that it generates a sky map of 
large-scale anisotropies in the CMB temperature map, thus allowing a 
possible additional window into their causes.}
{Using this indicator, we find statistically significant excess of 
large-scale anisotropy at well over the $95\%$ confidence level. 
This anisotropy defines a preferred direction in the CMB data. 
We discuss the statistical significance of this direction compared 
to Monte Carlo data obtained under the statistical isotropy hypothesis, 
and also compare it with other such axes recently reported in the literature. 
In addition we show that our findings are robust with respect to the 
details of the method used, so long as the statistical noise is kept 
under control; and they remain unchanged compared to
the foreground cleaning algorithms used in CMB maps.
We also find that the application of our method to the first and three-year 
WMAP data produces similar results.}
{}


\maketitle


\section{Introduction}

Recent years have witnessed a tremendous accumulation of
high-precision cosmological data. In particular, 
the combination of the first-year and, more recently, the three-year
high-precision data from the Wilkinson Microwave Anisotropy Probe
(WMAP) (Bennett et al.~\cite{WMAPa,WMAPb}; Hinshaw et al.~\cite{Hinshaw06}) 
has produced a wealth of information about the early universe.
These data have provided evidence of a nearly spatially-flat universe 
with a primordial spectrum of almost scale-invariant density
perturbations, as predicted by generic inflationary models.

On large scales, however, a number of potentially important anomalous
features in the cosmic microwave background (CMB) have been reported
(see Copi et al.~\cite{Copi05,Copi06} for a detailed discussion),
including their low quadrupole and octopole moments 
(Spergel et al.~\cite{Spergel03},~\cite{Spergel06}),
the quadrupole-octopole axis alignment (Tegmark et al.~\cite{TOH03};
de Oliveira-Costa et al.~\cite{OT03}; Weeks~\cite{Weeks}; 
Bielewicz et al.~\cite{Bielewicz}; Wiaux et al.~\cite{Wiaux};
Abramo et al.~\cite{Abramo}), 
evidence of a North-South asymmetry (Hansen et al.~\cite{Hansen04a}; 
Eriksen et al.~\cite{Eriksen04a}; Bernui et al.~\cite{BVWLF}), 
indications of a preferred axis of symmetry around 
$(b=30^\circ, l=250^\circ)$ in galactic coordinates,%
\footnote{Throughout this paper we use galactic coordinates with 
equator defined by $b=90^\circ, \, l \in [0^\circ,360^\circ]$.} 
or directions of maximum asymmetry towards 
$(b=100^\circ, l=237^\circ)$ (Bunn \& Scott~\cite{Bunn-Scott00};
Copi et al.~\cite{Copi04}; Eriksen et al.~\cite{Eriksen04a};
Hansen et al.~\cite{Hansen04a,Hansen04b}; 
Wibig~\cite{WW}; Land \& Magueijo~\cite{Land-Magueijo05a,Land-Magueijo05b};
Schwarz et al.~\cite{Schwarz04}; Prunet, et al~\cite{Prunet};
Donoghue \& Donoghue~\cite{Donoghue}),
as well as indications of non-Gaussian features in the CMB temperature 
f\/luctuations (Copi et al.~\cite{Copi04}; Komatsu et al.~\cite{Komatsu};
Park~\cite{Park}; Chiang et al.~\cite{Chiang}; Vielva et al.~\cite{Vielva};
Eriksen et al.~\cite{Eriksen05}; Larson \& Wandelt~\cite{Larson};
Coles et al.~\cite{Coles}; Naselsky et al.~\cite{Naselsky}; 
Bernui et al.~\cite{BTV}). 

Clearly the study of such anomalies must take into account 
that they may have non-cosmological origins such as
unsubtracted foreground contamination and/or systematics (Schwarz et 
al.~\cite{Schwarz04}), unconsidered local effects like gravitational 
lensing (Vale~\cite{Vale}; see, however, Cooray \& Seto~\cite{Cooray-Seto}),
or other mechanisms to break statistical isotropy (Gordon et 
al.~\cite{Gordon}; Tomita~\cite{Tomita}).
They may also have an extra-galactic origin 
(Bunn \& Scott~\cite{Bunn-Scott00}; Tegmark et al.~\cite{TOH03};
Eriksen et al.~\cite{Eriksen04a}; Hansen et al.~\cite{Hansen04a};
Land \& Magueijo~\cite{Land-Magueijo05a,Land-Magueijo05b}).
If they turn out to have a cosmological origin, however, this could
have far-reaching consequences on our understanding of the universe 
and, in particular, on the standard inflationary picture that predicts
statistically isotropic CMB temperature fluctuation patterns and
Gaussianity. 

In view of this, a great deal of effort has recently gone into verifying 
the existence of such anomalies by employing several different 
statistical approaches. 
Apart from corroborating the existence of a large-scale anisotropy, using 
multiple indicators may be useful in determining their 
origins. In addition, different indicators can in principle provide 
information about the multiple types of anisotropy that may be present 
(Souradeep \& Hajian~\cite{Souradeep-Hajian04,Souradeep-Hajian05}; 
Hajian \& Souradeep~\cite{Hajian03,Hajian05}; Hajian et al.~\cite{Hajian04}).

Here we propose a new indicator, based on the angular pair-separation 
histogram (PASH) method (Bernui \& Villela~\cite{BernuiVillela05}), 
as a measure of large-scale anisotropy. 
An important feature of this indicator is that it
generates a sky map of large-scale anisotropies for a given CMB 
temperature fluctuation map. 
This level of directional detail may also provide a possible additional 
window into their causes.

The structure of the paper is as follows. In Sect.~\ref{aniso}
we introduce our anisotropy indicator. Section~\ref{appli}
contains the results of the applying our indicator to the
WMAP data, and f\/inally Sect.~\ref{conclusion} contains the
summary of our main results and conclusions.

\section{Anisotropy Indicator}
\label{aniso}

In this section we construct an indicator that could measure the 
departure of CMB temperature fluctuation patterns from statistical 
isotropy. 
The construction of this indicator involves a number of steps. 
To begin, we recall that in CMB studies the celestial 
sphere is discretized into a 
set of equally sized pixels, each with a temperature fluctuation value. 
The first step involves the ordering of the pixels by temperature.
The pixels are then separated into two sets, one with positive and 
the other with negative temperature fluctuations, and each set is
subdivided into a number of submaps, each consisting of an equal number of 
consecutive pixels.%
\footnote{We shall see below that, subject to certain 
reasonable constraints, the details of this subdivision 
will not change our results significantly.}
This ordering and subdivision of the CMB map is motivated by the fact 
that any anisotropy in the CMB temperature fluctuations implies the 
presence of uneven distribution of the pixels with similar temperatures. 

To quantify this distribution, the next step is to generate the 
pair angular-separation hist\-ograms (PASH's), which are obtained 
by counting the number of pairs of pixels in a given submap with
angular separation $\alpha$ lying within small sub-intervals
(bins), $J_i$, of $( 0,180^\circ]$, of length 
$\delta\alpha = 180^\circ / N_{\mbox{\footnotesize bins}}$, where 
$J_i = \left( \alpha_i - \frac{\delta \alpha}{2} \, , \, \alpha_i +
\frac{\delta \alpha}{2}  \right],~ 
i=1,2, \dots ,N_{\mbox{\footnotesize bins}}\;,$
with the bin centers at $\alpha_i=\,(i-\frac{1}{2})\,\delta \alpha\,$.
The PASH is then defined by the normalized function
\begin{equation}
\label{PASH} \Phi(\alpha_i)=\frac{2}{n(n-1)}\,\,\frac{1}{\delta
\alpha} \, \sum_{\alpha \in J_i} \eta(\alpha) \; ,
\end{equation}
where $n$ is the number of pixels in the submap, $\,\eta(\alpha)$ 
the number of pairs of pixels with separation $\alpha$, and where the 
normalization condition 
$\sum_{i=1}^{N_{\mbox{\footnotesize bins}}} \Phi(\alpha_i)\,\,\delta \alpha=1\;$ 
holds.
In this way the pixels in each submap are ordered in a histogram.
Having calculated the PASHs for the submaps, we average them for
each bin to obtain the mean PASH (MPASH) 
$\langle \,\Phi(\alpha_i) \,\rangle\,$.

To obtain a measure of anisotropy for an observational pixelized map, 
we compare the MPASH calculated in this way with the histogram expected 
from a statistically isotropic map. 
To obtain this map, we first use the expected number of pairs
$\eta_{\mbox{exp}}(\alpha)$, with angular separation $\alpha \in J_i$,
for a distribution of $n$ pixels in the sky sphere $S^2$.
This allows the normalized expected pair angular-separation histogram
(EPASH) to be written as 
\begin{equation}
\label{defEPASH}
\Phi_{\mbox{exp}}(\alpha_i)=\frac{1}{N}\,\,\frac{1}{\delta \alpha}\,
              \sum_{\alpha \in J_i} \eta_{\mbox{exp}}(\alpha) =
\frac{1}{\delta \alpha}\,\,{\cal P}_{\mbox{exp}}(\alpha_i) \; ,
\end{equation}
where $N= n (n-1)/2$ is the total number of pairs of pixels, 
${\cal P}_{\mbox{exp}}(\alpha_i)=\sum_{\alpha \in J_i} \eta_{\mbox{exp}}(\alpha) / N$ 
is the expected probability that a pair of objects can be separated by an 
angular distance that lies in the interval $J_i$, and where the coefficient 
of the summation is a normalization factor. Defined in this way, 
$\Phi_{\mbox{exp}}(\alpha_i)$ gives the probability distribution
for finding pairs of points (pixels in a submap)
on the sky sphere with angular separations $\alpha_i \in (0, 180^\circ]$.%
\footnote{In this connection it is instructive to note that, for a 
homogeneous distribution of points on $S^2$, all angular separations 
$\,0 < \gamma \leq 180^\circ\,$ are allowed, and the corresponding 
probability distribution can be calculated to give 
${\cal P}^{\mbox{\footnotesize full-sky}}_{\mbox{exp}}(\gamma) 
= \frac{1}{2}\,\sin \gamma\,$.
This represents the limit of a statistically isotropic distribution 
of points in $S^2$ as the number of points go to infinity. 
One can thus quantify departure from statistical isotropy by 
calculating the departure of the mean observed probability distribution 
$\langle \,\Phi_{obs}(\alpha_i)\,\rangle$ from this quantity, i.e. by 
evaluating 
$\langle \,\Phi_{obs}(\alpha_i)\, \rangle - \Phi_{exp}(\alpha_i)\,$.}

We denote the difference between the MPASH, 
$\langle \, \Phi_{\mbox{obs}}(\alpha_i)
\,\rangle $, calculated from the pixelized observational map, and the 
EPASH $\,\Phi_{\mbox{exp}}(\alpha_i)\,$, obtained from an statistically 
isotropic map, as 
\begin{equation} \label{AnisInd1}
\Upsilon (\alpha_i) \equiv \langle \, \Phi_{\mbox{obs}}(\alpha_i) \,\rangle -
\,\Phi_{\mbox{exp}}(\alpha_i) \,\;.
\end{equation}
The latter can be obtained from an analytical expression 
derived by Teixeira (\cite{Teixeira03}), which gives the expected 
probability that an arbitrary pair of pixels, in a spherical cap of 
aperture $\gamma_0$ degrees, 
is separated by an angle $\gamma$ [$\,\gamma \in (0, 2\gamma_0 \leq 180^\circ\,$)]
\begin{eqnarray} \label{EPASH} 
{\cal P}_{\mbox{exp}}(\gamma; \gamma_0) \, = \, 
\frac{\sin \gamma}{\pi\,(1-\cos \gamma_0)^2} \,
\left( \, \arcsin [\,\{(\cos \gamma_0 + \cos \gamma) \right.   
&& \nonumber \\
\sqrt{\cos \gamma - \cos 2\gamma_0}\, \} \,
/ \, \{ \sin \gamma_0\, (1 + \cos \gamma_0)\,\sqrt{1+\cos \gamma} 
\, \} \,] && \nonumber  \\
\left. + \,\,(1 - 2\,\cos \gamma_0)\,
\arccos[\, \cot \gamma_0\, \tan (\gamma/2) \,]\, 
\right) \,. &&  
\end{eqnarray}
Alternatively, one can approximate $\Phi_{\mbox{exp}}$ by 
the MPASH, $\langle \, \Phi \,\rangle$, obtained through the 
average of the PASH's produced from a set of Monte Carlo--generated 
statistically isotropic maps (see Bernui \& Villela~\cite{BernuiVillela05} 
for details). 

\begin{figure}
\includegraphics[width=8.8cm,height=5.1cm]{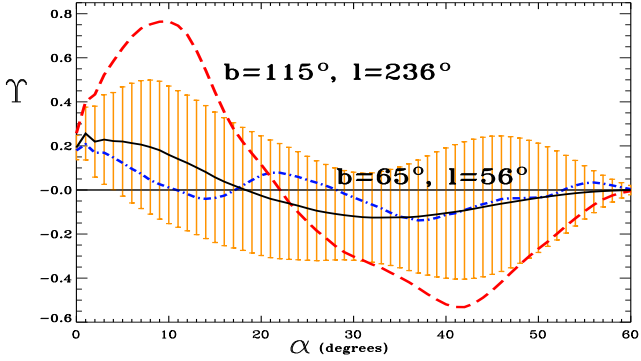}
\caption{\label{F1}  Plots of the difference histogram
$\Upsilon(\alpha)$
illustrating the magnitudes of the deviation from isotropy as a function 
of separation angle calculated for two specific spherical caps, centered 
at $(b=65^\circ, l=56^\circ)$ and $(b=115^\circ, l=236^\circ)$, 
with a $\gamma_0=30^\circ$ of aperture.
Plotted also, as a continuous line, is
the mean $\Upsilon$ corresponding to the set 
of spherical caps into which the sphere was subdivided, along with
the corresponding two-standard deviation region as a shaded area. 
As can be seen, the measured anisotropy is significant in one cap 
[$(b=115^\circ, l=236^\circ)$] but not in the other.
}
\end{figure}

Clearly, the difference histogram $\Upsilon$ gives the 
departure of the mean angular-separation probability distribution, 
obtained from observational data 
$\langle\,\Phi_{\mbox{obs}} \,\rangle$, 
with respect to the probability distribution of the statistically 
isotropic case. 
In this way, it contains information about the amplitude and 
angular scale of the large-angle correlations for the sky sphere, but 
no directional information about possible anisotropies. 
However, $\Upsilon $ can also be calculated for sections of 
the sphere, in which case it can be used to study the differences in 
the extent of anisotropy in different regions of the sky.
As an illustration of this point, we have plotted in Fig.~\ref{F1} 
difference histograms $\Upsilon $ corresponding to two 
different (antipodal) spherical caps centered at 
$(b=65^\circ, l=56^\circ)$ and $(b=115^\circ, l=236^\circ)$, 
each with an aperture of $\gamma_0=30^\circ$.
These histograms illustrate the magnitude of the deviation from
isotropy as a function of separation angle. 
As can be seen, the large-scale angular correlations are significant 
in one cap but not in the other, 
in comparison with the 2-standard deviations error bars. 

The study of the individual caps can thus provide hints
of possible large-scale anisotropies. But for a more systematic
study, one needs to understand how the anisotropy varies 
directionally over the whole sphere.
One therefore requires an indicator that encodes
a measure of anisotropy as a scalar function defined over 
the celestial sphere, rather than a histogram for each cap.

To define this indicator, let 
$\Omega_j \equiv \Omega(\theta_j,\phi_j;\gamma_0) \in
S^2$ be a spherical cap, with an aperture of $\gamma_0$ degrees,
centered at $(\theta_j,\phi_j)$, for $j=1, \ldots, N_{\mbox{caps}}$, 
where the union of the $N_{\mbox{caps}}$ caps covers 
the whole celestial sphere $S^2$. 
Define the scalar function $\sigma: \Omega_j \mapsto \cal{R}^{+}$, that 
assigns to the $j^{\,\rm{th}}-$cap, centered at $(\theta_j,\phi_j)$, a real 
positive number $\sigma_j$ given by the variance of the $\Upsilon$
(which has zero mean)
\begin{equation} \label{AnisInd2}
\sigma^2_j  \equiv \frac{1}{N_{\mbox{bins}}}
\sum_{i=1}^{N_{\mbox{\footnotesize bins}}} \Upsilon^2_j (\alpha_i)\;,
\end{equation}
where $\Upsilon_j \equiv \Upsilon(\theta_j,\phi_j)$ is the
$\Upsilon$ for the cap centered at $(\theta_j,\phi_j)$.

To obtain a local directional indicator, we cover the celestial 
sphere with a number of evenly distributed spherical caps and then
calculate $\sigma_j$ for each cap.
The $\sigma_j$ obtained in this way for each cap can then be viewed
as a measure of anisotropy in the direction of the center of that cap. 
Patching together the $\sigma$ values for each cap, we obtain
the indicator $\sigma = \sigma(\theta,\phi)$ defined over the
whole celestial sphere, which measures the deviation from isotropy
as a function of direction $(\theta,\phi)$. In this way, 
$\sigma(\theta,\phi)$ gives a scalar directional measure of anisotropy
over the celestial sphere. 

Now, since $\sigma (\theta,\phi)$ is a discrete scalar function 
defined on $S^2$, we can further quantify the details 
of its anisotropy content, by expanding it in spherical harmonics in the 
form
\begin{equation}
\sigma (\theta,\phi) = \sum_{\ell=0}^\infty \sum_{m=-\ell}^{\ell}
b_{\ell m} \,Y_{\ell m} (\theta,\phi) 
\end{equation}
and calculate the corresponding $\sigma$ power spectrum 
\begin{equation}
D_{\ell} = \frac{1}{2\ell+1} \sum_m |b_{\ell m}|^2 \; .
\end{equation}
It then follows that, if a large-scale asymmetry is present in the
original temperature distribution, it should significantly affect
the $\sigma$--map on the corresponding angular scales (i.e. its
lower multipoles).

To recapitulate the multiple steps involved in defining our
indicator $\sigma$, we itemize the procedure below:
\begin{enumerate}
\item[(i)] Generate a set of evenly distributed
overlapping spherical caps;
\item[(ii)] For each cap, divide the set of pixels into a number of submaps
$N_{\mbox{his}}$, each containing an equal number of pixels
within a temperature range;
\item[(iii)] Calculate the PASH, $\Phi_{\mbox{obs}}$, for each submap; 
\item[(iv)] Average over all the $\Phi_{\mbox{obs}}$
for each cap to obtain $\langle \Phi_{\mbox{obs}} \rangle$;
\item[(v)] Obtain the difference histogram $\Upsilon$ for each
cap by subtracting $\Phi_{\mbox{exp}} - \langle \Phi_{\mbox{obs}} \rangle$;
\item[(vi)] Calculate the $\sigma$ value, using Eq.~(5),
for each cap;
\item[(vii)] Define the $\sigma$--map as the discrete function %
$\sigma(\theta,\phi)$ on $S^{2}$ where $\theta$ and
$\phi$ are the coordinates of the center of each cap;
\item[(viii)] Calculate the multipoles $D_{\ell}$ for the $\sigma$--map.
\end{enumerate}

{}From the above construction of $\sigma(\theta,\phi)$ and its power spectrum, 
it is clear that there is no direct relation between the temperature 
multipoles and the $\sigma(\theta,\phi)$ multipoles. Although they are 
implicitly related, the relation is likely to be very complicated. 
In this work we take a more pragmatic approach by treating
our new anisotropy indicator as complementary to
the usual temperature-based indicators, without attempting 
to establish a direct connection between them.

In the next section we shall apply the indicator $\sigma(\theta,\phi)$ 
to both the first and three-year WMAP data. 
  
\section{Application to WMAP data}
\label{appli}

The WMAP measurements have produced high angular resolution maps of the 
temperature field of the CMB. 
Here we considered both the first-year and the three-year WMAP data.
For the first-year data we employed the full-sky foreground-cleaned 
Lagrange-ILC (LILC) map (Eriksen et al.~\cite{Eriksen04b}), the TOH map 
(Tegmark et al.~\cite{TOH03}), as well as the coadded WMAP map (Hinshaw 
et al.~\cite{Hinshaw03}).
In all cases we chose HEALPix parameter $N_{\mbox{side}}=128$
(G\'orski et al.~\cite{Gorski}), 
corresponding to a partition of the celestial sphere into 196608 pixels.
For the three-year WMAP data we used the coadded map and the 
foreground-cleaned maps ILC (Hinshaw et al.~\cite{Hinshaw06}), 
OT (de Oliveira-Costa \& Tegmark~\cite{OT06}), and PPG (Park et al.~\cite{PPG06}).

In all, we performed the $\sigma$--map analysis for seven CMB 
temperature maps: three using the first-year data and four with the 
full three-year data, all with different foreground cleaning algorithms. 
However, since the results are largely similar, we only present a 
detailed analysis of the three-year WMAP ILC map in the following. 
We also present the $\sigma$--maps for four of the remaining maps 
for comparison (the other two are similar to the depicted maps and were
dropped to avoid repetition).

In our calculations of the $\sigma$--map for these CMB maps, we
subdivided the sphere into 
a set of $N_{\mbox{caps}}=12288$ spherical caps of radius $30^{\circ}$ 
co-centered with the same number of pixels generated by HEALPix with 
$N_{\mbox{side}}=32$. 
We subdivided each cap into $N_{\mbox{his}}=64$ submaps, each consisting of the 
same number of pixels, and the number of bins chosen to produce the PASHs was 
taken to be $N_{\mbox{bins}}=180$. 

\begin{figure}
\includegraphics[width=8.8cm,height=5.1cm]{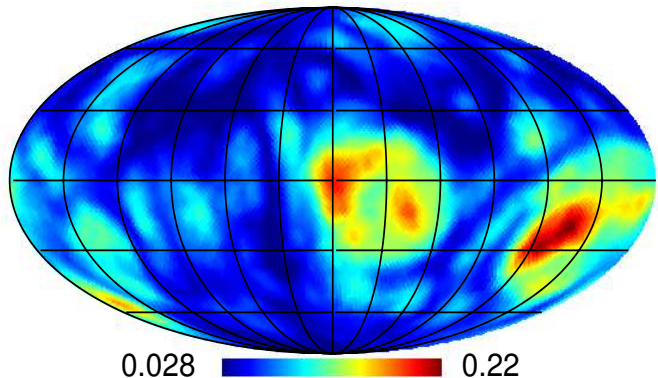}
\caption{\label{F2} The full sky map of $\sigma(\theta,\phi)$
calculated from 12288 evenly distributed $30^\circ$ caps covering the
ILC map.} 
\end{figure}

Figure~\ref{F2} shows the Mollweide projection of the $\sigma$--map in 
galactic coordinates obtained from the ILC WMAP map. 
As can be seen, the $\sigma$--map shows a high--$\sigma$
region suggestive of a  dipole-like distribution.
This result indicates that the distribution of hot and cold  
temperature pixels are not evenly distributed in the celestial sphere.
We note that this is different from a temperature dipole, where one half 
of the sky would, on average, be hotter than the other. 
An anisotropy $\sigma$-dipole is not directly related to the temperature 
dipole, which was removed in all the CMB maps we have used, but is instead  
a result of large-angle anisotropies that are not readily apparent from a 
simple inspection of the temperature map alone. 
Particularly prominent is a very high--$\sigma$ region concentrated on 
the south eastern corner of the $\sigma$--map with a well-defined 
maximum at $(b = 108^\circ, l = 222^\circ)$. 
This is within a $16.6^{\circ}$ separation from the directions recently 
indicated by Eriksen et al.~(\cite{Eriksen04a}) and Land \& 
Magueijo~(\cite{Land-Magueijo05a}).

\begin{figure}
\includegraphics[width=8.8cm,height=5.1cm]{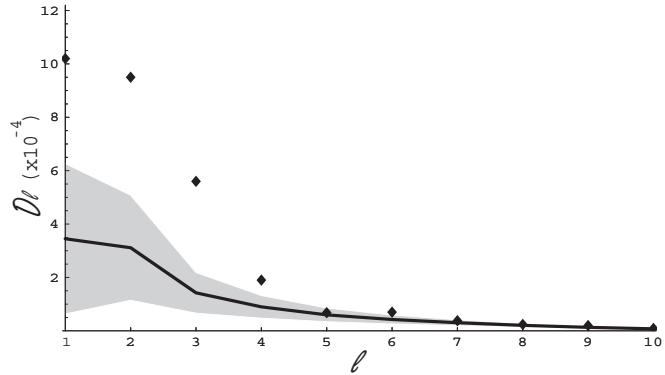}
\caption{\label{F3} The power spectrum of the  $\sigma$--map for
the ILC map, for $\ell=1, \cdots ,10$ (solid dots). 
Note the high values of the first three multipoles.
For comparison, the  mean expected power spectrum of the $\sigma$--map
obtained by averaging over a set of 1000 Monte Carlo-generated, 
statistically isotropic CMB maps (solid curve). Shown also are
the corresponding cosmic variance bounds for $\sigma$--maps
(shaded area).}
\end{figure}

To obtain more quantitative information about the observed anisotropy, 
we calculated the power spectrum of the $\sigma$--map for the ILC map, 
which is represented by solid dots in Fig.~\ref{F3}.
%
%
To estimate the statistical significance of these $\sigma$--multipole
values, we compared the power spectrum of the $\sigma$--map, calculated for 
the ILC map, with the averaged power spectrum of the $\sigma$--map obtained 
by averaging over a set of 1,000 Monte Carlo-generated statistically 
isotropic CMB maps.
Considering the WMAP best-fitting angular power spectrum of the $\Lambda$CDM
model (Spergel et al.~\cite{Spergel06}), each Monte Carlo CMB map is a stochastic 
realization of this spectrum obtained through randomized multipole components 
$a_{\ell m}$ generated within the cosmic variance limits.

The $\ell$ values used in this comparison lie in the range $1 \le \ell \le 10$,
which are sufficient to allow the investigation of the large-scale 
angular correlations corresponding to angular separations 
$\gtrsim 18^{\circ}$ 
in the $\sigma$--map. 
Moreover, we observe that $D_{\ell} \simeq 0$ for $\ell \geq 8$.
Averaging over the 1,000 spectra of the $\sigma$--maps obtained from the 
Monte Carlo CMB maps gives the expected power spectrum,  
which is depicted in Fig.~\ref{F3} (continuous line) along with the 
corresponding cosmic variance\footnote%
{We have checked numerically that the $b_{lm}$s of the $\sigma$--maps 
derived from the Monte Carlo CMB maps do follow Gaussian  distributions, 
which justifies the use of cosmic variance bounds and the $\chi^2$ test.}
(shaded area). 

We note that the very high dipole value (corresponding to $\ell=1$) found 
here is consistent with the $\sigma$--map (see Fig.~\ref{F2}) showing a 
clear separation of the higher and lower values for $\sigma$ into roughly 
two hemispherical regions. 
Using a $\chi^2$ test we have found this high value of the dipole 
to be statistically significant at $97\%$ confidence level, compared to 
the expected isotropic power spectrum. 
Note also the high values of the quadrupole and octopole moments that 
were found to be significant at $99\%$ and over $99.9\%$ (corresponding 
to a likelihood of one part in $10^4$ in a statistically isotropic map), 
respectively.
We also find that the likelihood of the first three multipoles of an 
isotropically-generated map to have such high values simultaneously is 
very small being less than one part in $10^5$.
Of the subsequent multipoles, only $D_4$ and $D_6$ fall outside the 
cosmic variance bounds.

\begin{figure*}
\includegraphics[width=8.8cm, height=8.5cm]{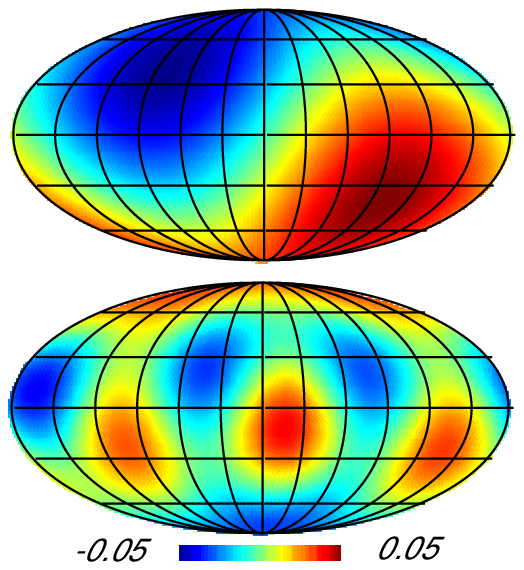}
\includegraphics[width=8.8cm, height=8.5cm]{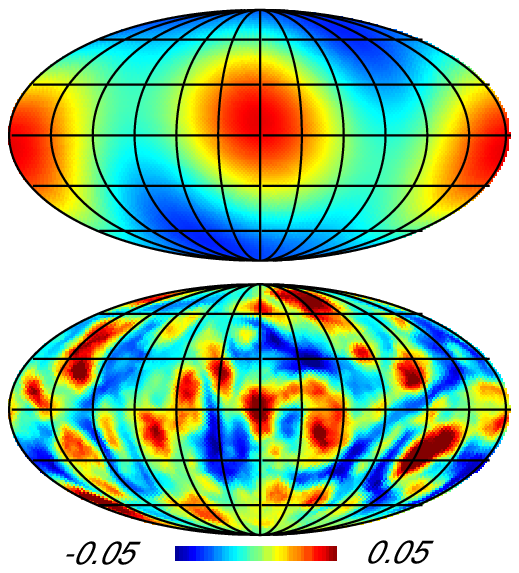}
\caption{\label{F4} Depicted clockwise are the ILC $\sigma$--map's
dipole, quadrupole, octopole, and  the remaining $\sigma$--map 
from which all these components plus the monopole were removed.}
\end{figure*}

\begin{figure*}
\includegraphics[width=8.8cm,height=9cm]{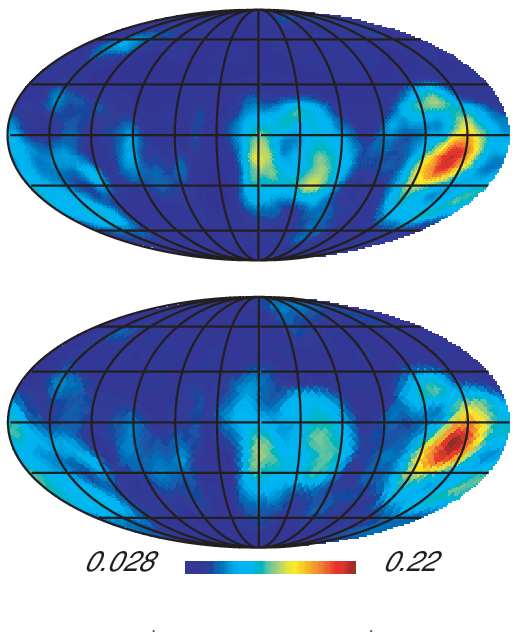}
\includegraphics[width=8.8cm,height=9cm]{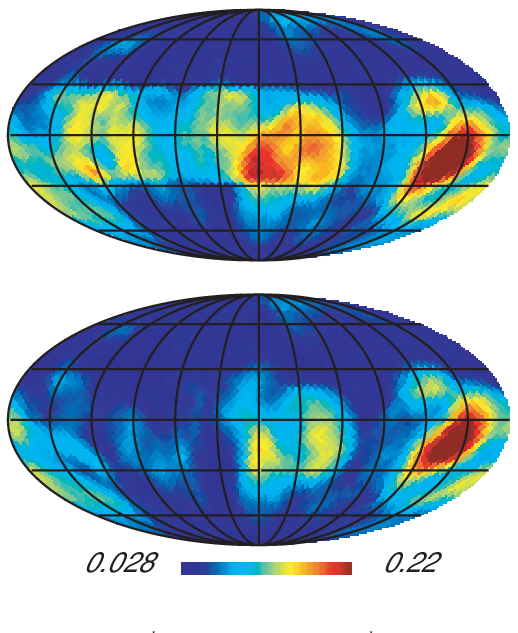}
\caption{\label{F5} Depicted clockwise are the $\sigma$--maps: 
LILC, three-year coadded map (showing the effects of the unfiltered 
galactic contamination), OT and TOH WMAP maps. 
The one-year coadded and the PPG maps are not shown as they are 
very similar to the three-year coadded and the ILC maps, respectively.
Note the high--$\sigma$ region in the south eastern corner of the 
$\sigma$--maps, which remains robust for all maps.}
\end{figure*}

To check the robustness of our results we also
calculated the power spectrum $\sigma$--map
obtained from the WMAP PPG map. We found that the first
three multipoles remain high relative to
what would be expected from an isotropic case,
at above $97\%$ statistical significance.

Given the anisotropic nature of the calculated $\sigma$--map, it is worth 
analyzing the shape of the anomalous multipoles in more detail. 
To this end, we depict in Fig.~\ref{F4} the dipole, the quadrupole, 
and the octopole, as well as the full $\sigma$--map with these three 
components and the monopole removed. 
The analysis of the dipole alone gives a direction towards 
$(b=130^\circ, l=259^\circ)$, which on its own does not agree very well
with the axis of maximum asymmetry found by Eriksen et al.~(\cite{Eriksen04a}) 
and also by Land \& Magueijo~(\cite{Land-Magueijo05a}). 
However, the sum of the first three multipoles has two maxima, one 
close to the galactic center, at $(b=97^\circ, l=342^\circ)$ and probably 
due to unremoved foregrounds, and the other in the direction 
$(b=112^\circ, l=217^\circ)$, which is in much better 
agreement with the axis of maximum asymmetry indicated 
by Eriksen et al.~(\cite{Eriksen04a}) and Land \& 
Magueijo~(\cite{Land-Magueijo05a}).

The quadrupole component has a very peculiar shape. 
It has two maxima and two minima lying symmetrically along a great circle.
To determine this feature more precisely, we used the method proposed 
by de Oliveira-Costa et al.~(\cite{OT03}), where for each given multipole 
(in their case, of the temperature map) the direction for which `angular 
momentum dispersion' is maximized is said to be a preferred direction. 
Applying this to the $\sigma$--map, we find that the angular momentum 
dispersion for the quadrupole component  is maximized in the direction 
$(b=80^\circ,\, l=316^\circ )$. 

The calculation of our anisotropy indicator requires not only 
the choice of a CMB map as input, but also the specification of a 
number of ingredients and parameters whose choice could in principle 
affect the outcome of our results. 
To test the robustness of our scheme, hence of our results, we 
studied the effects of changing various parameters employed in the 
calculation of our indicator. 
We found that so long as the statistical noise (see Bernui and 
Villela~\cite{BernuiVillela05}) satisfies
\begin{equation} 
\frac{\sqrt{N_{\mbox{bins}}}}{n \sqrt{N_{\mbox{his}}}} \lesssim 0.05,
\end{equation}
where $N_{\mbox{his}}$ is the number of submaps and $n$ is the mean 
number of pixels in each submap, then the $\sigma$--map and its power 
spectrum do not change appreciably as we change the number of submaps 
$N_{\mbox{his}}$ from 2 to 64, the size of the spherical caps between 
$15^{\circ}$ and $30^{\circ}$, the number of spherical caps, $N_{\mbox{caps}}$
(with values 768, 3072, and 12288, corresponding
to HEALPix parameter $N_{\mbox{side}}=8, 16$, and $32$, respectively)
and the $N_{\mbox{bins}}$ values
in the range 180 to 400.

Given that galactic foregrounds are likely to leave their most 
pronounced effects in the neighborhood of the galactic equator, 
one may suspect 
that the pronounced features, such as the high $\sigma$ region
in the south eastern corner of the $\sigma$--map 
(see Fig. (\ref{F2})),
may be an artifact of the foreground contamination processes.

To show that this not the case, we first note that the latter 
feature lies outside the exclusion region defined by the Kp0 mask,
which is the severest  cut-sky WMAP mask. 
To proceed, we also calculated the $\sigma$--maps for the other 6 CMB 
temperature maps. 
We studied the three-year (Hinshaw et al.~\cite{Hinshaw06})
and one-year coadded (Hinshaw et al.~\cite{Hinshaw03}) maps, 
which are the combination of eight (from Q1 trough W4) foreground-reduced 
sky-maps in the Q--, V--, and W--bands, and 
minimized but did not seek to eliminate the galactic contamination.
We also studied the foreground-cleaned TOH and OT maps given by Tegmark 
et al.~(\cite{TOH03}) and de Oliveira-Costa \& Tegmark~(\cite{OT06}),
as well as the LILC (Eriksen et al.~\cite{Eriksen04b})
and PPG (Park et al.~\cite{PPG06}) 
CMB maps. 
The corresponding $\sigma$--maps are shown in Fig.~\ref{F5} (the PPG and 
one-year coadded maps were dropped to avoid repetition).
Together, these figures show that, despite differences in detail, the prominent 
high--$\sigma$ region in the south eastern corner remains robust. 

These results indicate that the main anisotropic feature we have 
found is independent of the foreground cleaning processes employed.
In addition we recently (Mota et al.~\cite{MBRT2007}) performed a 
similar analysis of the WMAP one-year LILC data with both Kp0 and Kp2 
masks applied and found very similar results.
We also showed that this feature remains essentially the same 
in the Q, V and W bands (when analyzed after applying the Kp2 mask)  
which would not be expected if it were an artifact of uncleaned galactic 
foreground. 

We should also note that apart from the high $\sigma$ region in the 
south eastern corner, which lies outside the Kp0 mask,
there is another weaker structure around the galactic center. 
Although this structure remains in some form in all the maps we have 
analyzed its intensity and other details vary considerably for different choices
of the cleaning algorithms, as can be seen from Figs.~\ref{F2} and~\ref{F5}.
Considering that it lies in the most contaminated region of the galactic 
plane, well inside the Kp0 mask, it is not clear whether this feature is 
not the result of the galactic foreground contamination. Furthermore its 
contributions to the excess dipole, quadrupole, and octopole is relatively 
minor compared to the high--$\sigma$ region in the south-eastern corner.

\section{Conclusions}
\label{conclusion}

We have proposed a new method of measuring directional deviations   
from statistical isotropy in the CMB sky, in order to study the 
possible presence and nature of large-scale anisotropy in the WMAP 
data. 

Our anisotropy indicator has enabled us to construct a 
$\sigma$--map in order to search for evidence of 
large-angle anisotropies in the WMAP CMB temperature field. 
In particular we have found, with high statistical significance, 
a small region in the  celestial sphere with very high values 
of $\sigma$, indicative of an axis of asymmetry that defines a direction 
close to the one reported recently (Eriksen et al.~\cite{Eriksen04a}; 
Land \& Magueijo~\cite{Land-Magueijo05a}).

To obtain a more quantitative measure of this anisotropy, we studied
the spherical harmonic expansion of the $\sigma$--map generated from
the ILC map and found that the corresponding dipole component is 
significantly higher than would be expected, as compared with an average 
of a set of $\sigma$--maps obtained from 1000 Monte Carlo CMB maps 
generated under the statistical isotropy hypothesis. 
Moreover, we have found that the combination of the dipole, quadrupole,
and octopole components of the $\sigma$--map has a strongly preferred
direction close to the small region of the maximum value of $\sigma$.

Using a $\chi^2$ test we  found that this high value of the 
dipole is statistically significant at a $97\%$ confidence level (CL), 
compared to the expected isotropic case. 
We also found high values of the quadrupole and octopole moments,
which were found to be significant at $99\%$ CL and over 
$99.9\%$ CL (corresponding to a likelihood of one part in 
$10^4$ for statistically isotropic generated maps) respectively.
The likelihood of the first three multipoles
of an isotropically generated map to have such high values simultaneously 
is very small at less than one part in $10^5$.

We have shown that the results reported here are robust by showing 
that the $\sigma$--map does not significantly change with different
choices of parameters,
so long as the statistical noise is kept under control.
We have also studied the ef\/fects of dif\/ferent foreground-cleaning
algorithms, or absence thereof, by considering the LILC (first-year), 
TOH (first-year), OT (three-years), PPG (three-year), and the coadded 
(first and three-year) CMB maps in addition to WMAP three-year
ILC map.
 
We have found again that the corresponding $\sigma$--maps remain qualitatively 
unchanged, in that the high--$\sigma$ region in the south eastern corner of the 
$\sigma$--map remains essentially invariant for all the maps considered 
here, while similar statistical significances for the most important 
(low-$\ell$) multipoles $D_\ell$ hold for all CMB maps considered.
This provide evidence that our results regarding the large-angle 
anisotropy in the CMB maps remain unchanged no matter whether the 
first or three-year data are used.

This robustness demonstrates that our indicator is well-suited to 
the study of anisotropies in the CMB data. Furthermore, although 
derived from the temperature maps, it reveals features 
that are not apparent in the analysis of the temperature maps alone. 
Our scheme for studying directional anisotropies is complementary to the 
existing approaches in the literature, in the sense that it reveals
different aspects of the same data set. 
However, the relationship between them is clearly non-trivial and requires
further investigation, which is beyond the scope of the present work.

Regarding the origin of such large-angle anisotropy, a number of 
suggestions have been put forward. 
Briefly, they can arise either from a subtle form of unremoved 
foreground contamination (in which case the $\sigma$--map might indicate 
where in the sky this contamination is most intense) or from the 
universe being genuinely anisotropic on large angular scales. 
This possibility is particularly interesting, as it would have 
potentially important consequences for the standard inflationary picture, 
which predicts statistically isotropic CMB temperature fluctuation 
patterns. 

Among this type of suggestions, it has been proposed that the
preferred direction could be due to the universe possessing a 
non-trivial topology (see, e.g., Copi et al.~\cite{Copi05}; 
Spergel et al.~\cite{Spergel03}; 
for more details on cosmic topology see, e.g., the review articles 
Lachi\`{e}ze-Rey \& Luminet~\cite{CosmTopReviews};
Starkman~\cite{Starkman}; Levin~\cite{Levin}; 
Rebou\c{c}as \& Gomero~\cite{RG}; 
Lehoucq et al.~\cite{TopSign}; 
Roukema \& Edge~\cite{Roukema};
Cornish et al.~\cite{Cornish};
Bond et al.~\cite{Bond00a,Bond00b};
Fagundes \& Gausmann~\cite{Fagundes}; 
Uzan et al.~\cite{Uzan}; 
Gomero et al.~\cite{Gomero00,Gomero01a,Gomero01b,Gomero01c,Gomero02}; 
Mota et al.~\cite{Mota03,Mota04};
Luminet et al.~\cite{Luminet03}; 
Aurich et al.~\cite{Aurich05a,Aurich05b}; 
Hipolito-Ricaldi \& Gomero~\cite{Hipolito};
Weeks et al.~\cite{Weeks03a};
Weeks~\cite{Weeks03b}). 
If topology is indeed the origin, our anisotropic indicator is 
promising for distinguishing between different topologies. 
On the other hand, the extent to which this is true for 
the $\sigma$-indicator is unclear and deserves further 
investigation.

Whatever the origin of these large-angle anomalies may be, the 
robustness of our anisotropy indicator seems to be sufficiently 
sensitive to reliably map them, which in turn could facilitate the 
task of explaining their origin.

\begin{acknowledgements}
We acknowledge use of the Legacy Archive for Microwave Background Data 
Analysis (LAMBDA), and  of the TOH \& OT maps 
the LILC map 
and the PPG map 
Some of the results in this paper were derived using the HEALPix package. 
We thank CNPq, PCI-CBPF/CNPq, PCI-INPE/CNPq, and PPARC for the grants 
with which this work was carried out. 
MJR thanks Glenn Starkman for fruitful discussions related to statistical 
isotropy.  
We are grateful to C.A. Wuensche for his advice on computer simulations. 
We also thank K. Land and T. Villela for useful comments. 
\end{acknowledgements}



\begin{thebibliography}{}

\bibitem[2006]{Abramo} Abramo, L. R., Bernui, A., Ferreira, I., Villela, T.,
\& Wuensche, C. A. 2006, \prd, 74, 063506

\bibitem[2005a]{Aurich05a} Aurich, R., Lustig, S., \& Steiner, F. 2005a, Class. Quantum Grav., 22, 2061

\bibitem[2005b]{Aurich05b} Aurich, R., Lustig, S., \& Steiner, F. 2005b, Class. Quantum Grav., 22, 3443  

\bibitem[2003a]{WMAPa} Bennett, C. L., Halpern, M., Hinshaw, G., et al. 2003a, \apjs, 148, 1   

\bibitem[2003b]{WMAPb} Bennett, C. L., Bay, M., Halpern, M., et al. 2003b, \apj, 583, 1

\bibitem[2006]{BTV} Bernui, A., Tsallis, C., \& Villela, T. 2006, Phys. Lett. A, 356, 426

\bibitem[2006]{BernuiVillela05} Bernui, A. \& Villela, T. 2006, \aap, 445, 795

\bibitem[2006]{BVWLF} Bernui, A., Villela, T., Wuensche, C. A., Leonardi, R., 
\& Ferreira, I. 2006, \aap, 454, 409 

\bibitem[2005]{Bielewicz} Bielewicz, P., Eriksen, H. K., Banday, A. J., 
G\'orski, K. M., \& Lilje, P. B. 2005, astro-ph/0507186

\bibitem[2000a]{Bond00a} Bond, J. R., Pogosyan, D., \& Souradeep, T. 2000a, \prd, 62, 043005  

\bibitem[2000b]{Bond00b} Bond, J. R., Pogosyan, D., \& Souradeep, T. 2000b, \prd, 62, 043006  

\bibitem[2000]{Bunn-Scott00} Bunn, E. F. \& Scott, D. 2000, \mnras, 313, 331

\bibitem[2003]{Chiang} Chiang, L.-Y., Naselsky, P. D., Verkhodanov, O. V., 
\& Way, M. J. 2003, \apj, 590, L65 

\bibitem[2004]{Coles} Coles, P., Dineen, P., Earl, J., \& Wright, D. 2004, 
\mnras, 350, 983 

\bibitem[2005]{Cooray-Seto} Cooray, A. \& Seto, N. 2005, astro-ph/0509039

\bibitem[2004]{Copi04} Copi, C. J., Huterer, D., \& Starkman, G. D. 2004,
\prd, 70, 043515  

\bibitem[2005]{Copi05} Copi, C. J., Huterer, D., Schwarz, D. J., \& Starkman, G. D. 
2005, astro-ph/0508047 

\bibitem[2006]{Copi06} Copi, C. J., Huterer, D., Schwarz, D. J., \& Starkman, G. D.
2006, astro-ph/0605135

\bibitem[1998]{Cornish} Cornish, N.J., Spergel, D., \& Starkman, G. 1998, 
Class. Quantum Grav., 15, 2657 

\bibitem[2003]{OT03} de Oliveira-Costa, A., Tegmark, M., Zaldarriaga, M.,
\& Hamilton, A. 2004, \prd, 69, 063516   

\bibitem[2006]{OT06} de Oliveira-Costa, A. \& Tegmark M. 2006,  
\prd, 74, 023005

\bibitem[2005]{Donoghue} Donoghue, E. P. \& Donoghue, J. F. 2005,
\prd, 71,  043002  

\bibitem[2004a]{Eriksen04a} Eriksen, H. K., Hansen, F. K., Banday, A. J., 
G\'orski, K. M., \& Lilje, P. B. 2004a, \apj, 605, 14 

\bibitem[2004b]{Eriksen04b} Eriksen, H. K., Banday, A. J., G\'orski, K. M., 
\& Lilje, P. B. 2004b, \apj, 612, 633 

\bibitem[2005]{Eriksen05} Eriksen, H. K., Banday, A. J., G\'orski, K. M., 
\& Lilje, P. B. 2005, \apj, 622, 58

\bibitem[1999]{Fagundes} Fagundes, H. V. \& Gausmann, E. 1999, Phys. Lett. A, 261, 235  

\bibitem[2000]{Gomero00} Gomero, G. I., Rebou\c{c}as, M. J., \& Teixeira, A. F. F. 2000, 
Phys. Lett. A, 275, 355

\bibitem[2001a]{Gomero01a} Gomero, G. I., Rebou\c{c}as, M. J., \& Teixeira, A. F. F. 2001a, 
Class. Quantum Grav., 18, 1885

\bibitem[2001b]{Gomero01b} Gomero, G. I., Rebou\c{c}as, M. J., \&  Tavakol, R. 2001b, 
Class. Quantum Grav., 18, 4461 

\bibitem[2001c]{Gomero01c} Gomero, G. I., Rebou\c{c}as, M. J., \&  Tavakol, R. 2001c,
Class. Quantum Grav., 18, L145 

\bibitem[2002]{Gomero02} Gomero, G. I., Teixeira, A. F. F., Rebou\c{c}as, M. J., 
\& Bernui, A. 2002, Int. J. Mod. Phys. D, 11, 869

\bibitem[2005]{Gordon} Gordon, C., Hu, W., Huterer, D., \& Crawford, T. 2005, 
astro-ph/0509301 

\bibitem[2005]{Gorski} G\'orski, K. M., Hivon, E., Banday, A. J., et al. 2005, 
\apj, 622, 759

\bibitem[2003]{Hajian03} Hajian, A. \& Souradeep, T. 2003, \apj, 597, L5
%
\bibitem[2004]{Hajian04} Hajian, A., Souradeep, T., \& Cornish, N. 2004, \apj, 618, L63 
%
\bibitem[2005]{Hajian05} Hajian, A. \& Souradeep, T. 2005, astro-ph/0501001

\bibitem[2004a]{Hansen04a} Hansen, F. K., Banday, A. J., \& G\'orski, K. M. 2004, 
\mnras, 354, 641

\bibitem[2004b]{Hansen04b} Hansen, F. K., Cabella, P., Marinucci, D., 
\& Vittorio, N. 2004, \apj, 607, L67 

\bibitem[2003]{Hinshaw03} Hinshaw, G., Spergel, D. N., Verde, L., et al. 
2003, \apjs, 148, 135

\bibitem[2006]{Hinshaw06} Hinshaw, G., Nolta, M. R., 
Bennett, C. L., et al. 2006, astro-ph/0603451  

\bibitem[2005]{Hipolito} Hipolito-Ricaldi, W. S. \& Gomero, G. I. 2005, 
\prd, 72, 103008 

\bibitem[2003]{Komatsu} Komatsu, E., Kogut, A., Nolta, M., et al. 2003, 
\apjs, 148, 119

\bibitem[1995]{CosmTopReviews} Lachi\`{e}ze-Rey, M. \& Luminet, J.-P. 1995, 
\physrep, 254, 135  

\bibitem[2005a]{Land-Magueijo05a} Land, K. \& Magueijo, J. 2005a, \mnras, 357, 994 

\bibitem[2005b]{Land-Magueijo05b} Land, K. \& Magueijo, J. 2005b, \prl, 95, 071301 

\bibitem[2004]{Larson} Larson, D. L. \& Wandelt, B. D. 2004, \apj, 613, L85  

\bibitem[1996]{TopSign} Lehoucq, R., Lachi\`{e}ze-Rey, M., \& Luminet, J.-P. 1996, 
\aap, 313, 339

\bibitem[2002]{Levin} Levin, J. 2002, \physrep, 365, 251  

\bibitem[2003]{Luminet03} Luminet, J.-P., Weeks, J. R., Riazuelo, A., Lehoucq, R., 
\& Uzan, J.-P. 2003, \nat, 425, 593

\bibitem[2004]{Mota04} Mota, B., Gomero, G. I., Rebou\c{c}as, M. J., \& Tavakol, R. 2004, 
Class. Quantum Grav., 21, 3361

\bibitem[2003]{Mota03} Mota, B., Rebou\c{c}as, M. J., \& Tavakol, R. 2003, 
Class. Quantum Grav., 20, 4837

\bibitem[2007]{MBRT2007} Mota, B., Bernui, A., Rebou\c{c}as, M. J., 
\& Tavakol, R. 2007, to appear in the \emph{Proceedings of the Second 
International Workshop on Astronomy and Relativistic Astrophysics},
Int.\ J.\ Mod.\ Phys.\ D, in press.

\bibitem[2004]{Naselsky} Naselsky, P. D., Chiang, L.-Y., Olesen, P., 
\& Verkhodanov, O. V. 2004, \apj, 615, 45 

\bibitem[2004]{Park} Park, C.-G. 2004, \mnras, 349, 313

\bibitem[2006]{PPG06} Park, C.-G., Park, C., \& Gott III, J. R. 2006, astro-ph/0608129

\bibitem[2005]{Prunet} Prunet, S., Uzan, J.-P., Bernardeau, F., \& Brunier, T. 2005,
\prd, 71, 083508

\bibitem[2004]{Souradeep-Hajian04} Souradeep, T. \& Hajian, A. 2004, Pramana, 62, 793 

\bibitem[2005]{Souradeep-Hajian05} Souradeep, T. \& Hajian, A. 2005, astro-ph/0502248

\bibitem[2003]{Teixeira03} Teixeira, A. F. F. 2003, physics/0312013

\bibitem[2005]{Tomita} Tomita, K. 2005, astro-ph/0509518

\bibitem[2004]{RG} Rebou\c{c}as, M. J. \& Gomero, G. I. 2004, Braz. J. Phys., 34, 1358

\bibitem[1997]{Roukema} Roukema B. F. \& Edge A. 1997, \mnras, 292, 105 

\bibitem[2004]{Schwarz04} Schwarz, D. J., Starkman, G. D., Huterer, D.,
\& Copi, C. J. 2004, \prl, 93, 221301 

\bibitem[2003]{Spergel03} Spergel, D. N., Verde, L., Peiris, H. V., et al. 2003, 
\apjs, 148, 175

\bibitem[2006]{Spergel06} Spergel, D. N., Bean, R., Dor\'e, O., et al. 2006, 
astro-ph/0603449

\bibitem[1998]{Starkman} Starkman, G. D. 1998, Class. Quantum Grav., 15, 2529

\bibitem[2003]{TOH03} Tegmark, M., de Oliveira-Costa, A., \& Hamilton, A. J. S. 
2003, \prd, 68, 123523

\bibitem[1999]{Uzan} Uzan, J.-P., Lehoucq, R., \& Luminet, J.-P. 1999, \aap, 351, 766 

\bibitem[2005]{Vale} Vale, C. 2005, astro-ph/0510137

\bibitem[2004]{Vielva} Vielva, P., Mart\'{\i}nez-Gonz\'alez, E., Barreiro, R. B., 
Sanz, J. L., \& Cayon, L. 2004, \apj, 609, 22 

\bibitem[2003a]{Weeks03a} Weeks, J. R., Lehoucq, R., \& Uzan, J.-P. 2003a, 
Class. Quantum Grav.,  20, 1529

\bibitem[2003b]{Weeks03b} Weeks, J. R. 2003b, Mod. Phys. Lett. A, 18, 2099

\bibitem[2004]{Weeks} Weeks, J. R. 2004, astro-ph/0412231

\bibitem[2005]{Wiaux} Wiaux, Y., Vielva, P., Mart\'{\i}nez-Gonz\'alez, E., \& 
Vandergheynst, P. 2006, \prl, 96, 151303, astro-ph/0603367

\bibitem[2005]{WW} Wibig, T. \& Wolfendale, A. W. 2005, \mnras, 360, 236

\end{thebibliography}
\end{document}